# YBa$_2$Cu$_3$O$_{6+\delta}$-originated attraction force giving rise to Meissner effect at 300 K

A. V. Fetisov and R. I. Gulyaeva

*Institute of Metallurgy, Ural Branch of the Russian Academy of Sciences, Ekaterinburg, Russia*

The exposure of YBCO to an air atmosphere at first at $p_{H_2O}$ = 110 Pa and then at $p_{H_2O}$ = 1 kPa has been used for uncharacteristic and unusual properties of YBCO to be observed. Among them there was an increased reactivity of YBCO with respect to water. Besides, it was experimentally obtained extremely excessive weight gain of hydrated YBCO samples that was not corresponding to the quantity of the absorbed water. Finally, we obtained the significant negative value of the room-temperature (RT) magnetic susceptibility in conventional magnetic study of the hydrated YBCO that was confirmed by the direct observation of decreasing weight of the same samples that were suspended above the source of controlled magnetic field (DC mode) at RT. All the results were explained on the basis of an approximate idea that there is a certain source of attractive force inside YBCO, acting on surrounding particles and bodies.

## 1. Introduction

It is now becoming clear that when describing the phenomenon of high-temperature superconductivity (HTSC), along with the traditional mechanism of electron-phonon coupling there is the need to consider the additional interactions between electrons (holes) for the high stability of the superconducting Cooper pairs with respect to temperature to be explained. The electron-exciton, magnetic and some others interactions are currently considered as possible options for that [1–3]. However, theories based on that have not given a satisfactory description of the HTSC phenomenon yet.

In the present work a series of experiments has been carried out allowing to consider that at least inside one representative of the HTSC family, namely, YBa$_2$Cu$_3$O$_{6+\delta}$ (YBCO) there is a certain source of attractive force (AF) acting on surrounding particles and bodies. AF intensity is expected to depend on the spatial configuration of the "source" and acted bodies (particles). Detection of AF is significantly easier when YBCO is under the influence of hydration with a low concentration of water vapor. In this case, its interaction with the surrounding bodies (including water molecules) is much magnified and it becomes not difficult to fix it with the standard methods. At room temperature (RT) a hydrated sample showed a significant negative value of the magnetic susceptibility and was ejected from permanent magnetic field that can be considered as a manifestation of the Meissner effect. Such growing $T_c$ is associated by us with increased intensity of AF in hydrated YBCO, suggesting that AF participates in the formation of Cooper pairs.

The present work is the logical completion of a long series of author's studies (e.g., [4–6]), in which various manifestations of the interaction of YBCO with the surrounding bodies were found, but the findings were not yet adequately explained.

## 2. Experiment details

YBCO was synthesized from oxides $Y_2O_3$ and CuO and barium carbonate at 940 °C for 150 h with intermediate regrinding and then a portion of the product (2–3 g) was saturated with oxygen at 585 °C (5 h) in flowing air. The sintered body was ground in an agate mortar and its metered portions were introduced into quartz micro-tubes with an internal diameter of 3 mm (the samples of type I, weighting 110±35 mg) or 7.7 mm (the samples of type II, weighting 250–800 mg). The tubes were covered with caps.

It should be noted that at all stages of high-temperature and oxidative processing we conducted a thorough purity control of the atmosphere, which the synthesized material interacted with. Immediately after the oxidative annealing, the material being still hot was placed inside a box with clean air where it was ground and packed. Finally, the prepared samples were exposed to a low humidity atmosphere ($p_{H_2O}$ = 110±10 Pa) for three days. The listed actions are an imperative for the sample ability to be hydrated at low $p_{H_2O}$ (probably due to the prepared in a special way surface). According to X-ray powder diffractometry, the product contained 96-97% main phase, which was referred to orthorhombic syngony with crystalline lattice parameters: $a$ = 3.8284 (1), $b$ = 3.8884 (1), and $c$ = 11.7063 (3). The average particle size of the YBCO powder was 20 μm. To control the oxygen content (6+δ) the iodometric titration technique was employed. The value obtained was 6.75±0.02.

Research of physical properties of YBCO was performed using a diffractometer Shimadzu XRD 7000 (Japan), a vibrating sample magnetometer Cryogenic CFS-9T-CVTI (UK), and a thermal analyzer Netzsct STA 449C Jupiter (Germany) combined with a mass spectrometer Netzsct QMS 403C Aëolos (Germany). In addition, an analytical balance Shimadzu AUW-120D was used as an important tool for the investigation. It allowed estimating the presence of new properties in the samples relatively easy – by analyzing the kinetics of the saturation of YBCO with air components occurring at low $p_{H_2O}$.

## 3. Results and discussion

With the aim of studying the adsorption properties of YBCO at room temperature (RT), dispersed samples of 110±15 mg (type I) were exposed to air (absolute humidity of about 1 kPa), being periodically weighted. When exposed, the samples were in vertical position (the importance of this information will clear up later). It is well known that the level of humidity used in the experiments is quite safe for YBCO prepared by a conventional method; the weight rise kinetics of this material is shown in Fig. 1 as curve *2*. Meanwhile the weight of our especially prepared samples was intensively growing under reported here conditions (see curve *1* in Fig. 1). Another bright feature of the tested powder was its agglomeration, already noticeable at the beginning stage of exposure to air. Herewith the strength of the interaction of powder particles with each other was fast-rising with increasing the duration of the sample exposure. At a weight increase of more than 1% the powder already was like a sintered body, withstanding pressure under the load of about 1.5 kgs·sm$^{-1}$. Such agglomeration is not typical even for substantially water-saturated ordinary YBCO.

X-ray spectra of original and saturated samples shown in Fig. 2 exhibit the main phase (YBCO) during the experiment to remain practically undestroyed, i.e., chemical degradation has took place in very limited volume.

To establish the nature of the gas adsorbed by YBCO, thermal analysis (TA) was carried out, the results of which are shown in Fig. 3a and Table 1.

According to the well-known thermodynamic properties of YBCO [7], the integrated thermal desorption represented in Fig. 3a by the thermogravimetric (TG) curves must include the escape of structural oxygen (temperature range 370–900 °C) as a constant part of the total gas release, which is about 1.4 wt.%. This part consists of: (a) oxygen contained in a sample before exposing to air; (b) oxygen absorbed by a sample in an adjacent temperature range 270–370 °C. Additional outgassing apparently relates to the air components absorbed at RT. As one can see in Fig. 3a and Table 1, corresponding weight loss data agree well with the amount of previously adsorbed gas. Shown at the bottom of Fig. 3a spectra relating to the desorbed water (mass spectral data are normalized to 100 mg of sample weight) permit to calculate the $H_2O$ proportion in the total gas exchange occurring at RT (see Table 1): 24±5 wt.%. The total proportion of other gaseous impurities, according to mass spectrometry, is about 2 wt.% (it is mainly $CO_2$). However, these impurities have also presented in the original sample and, therefore, they should not be considered as adsorbed at RT.

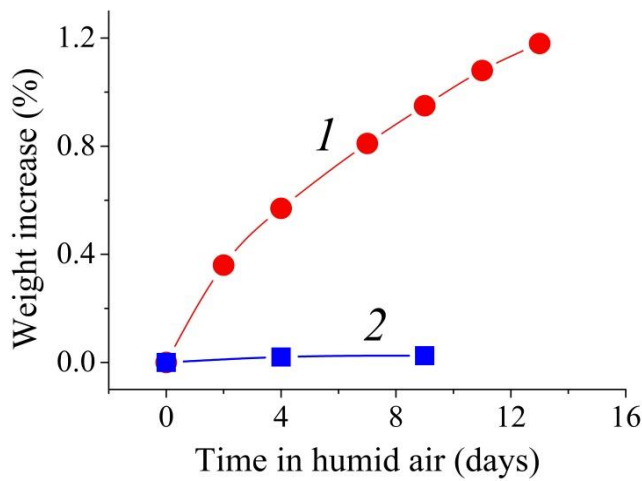

**Fig. 1.** The kinetic of the saturation of YBCO with air components. Curve *1* reflects the case when the material was pretreated in the dry atmosphere ($p_{H_2O}$ = 110 Pa). A kinetic curve obtained under the same conditions but for not treated YBCO is presented as curve *2*.

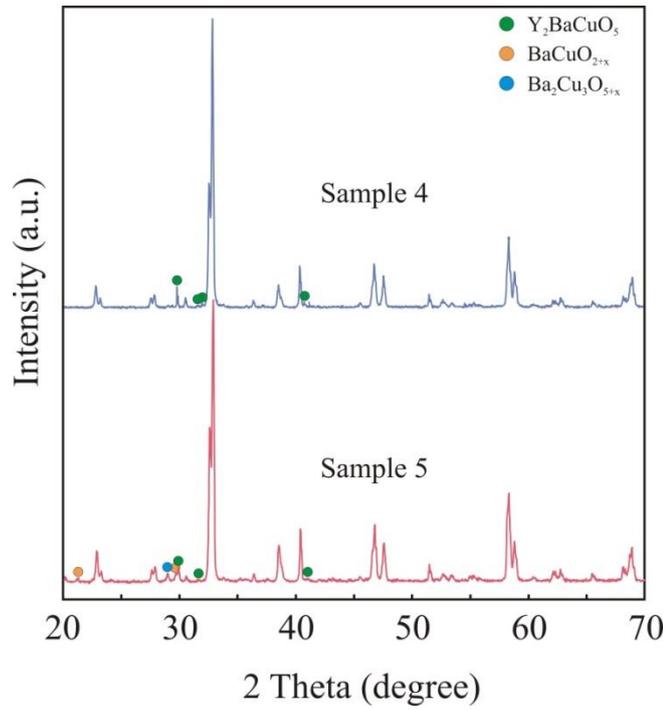

**Fig. 2.** X-ray spectra of the original (4) and hydrated (5) samples; the presence of the impurity phases (sample numbers are the same as in Table 1).

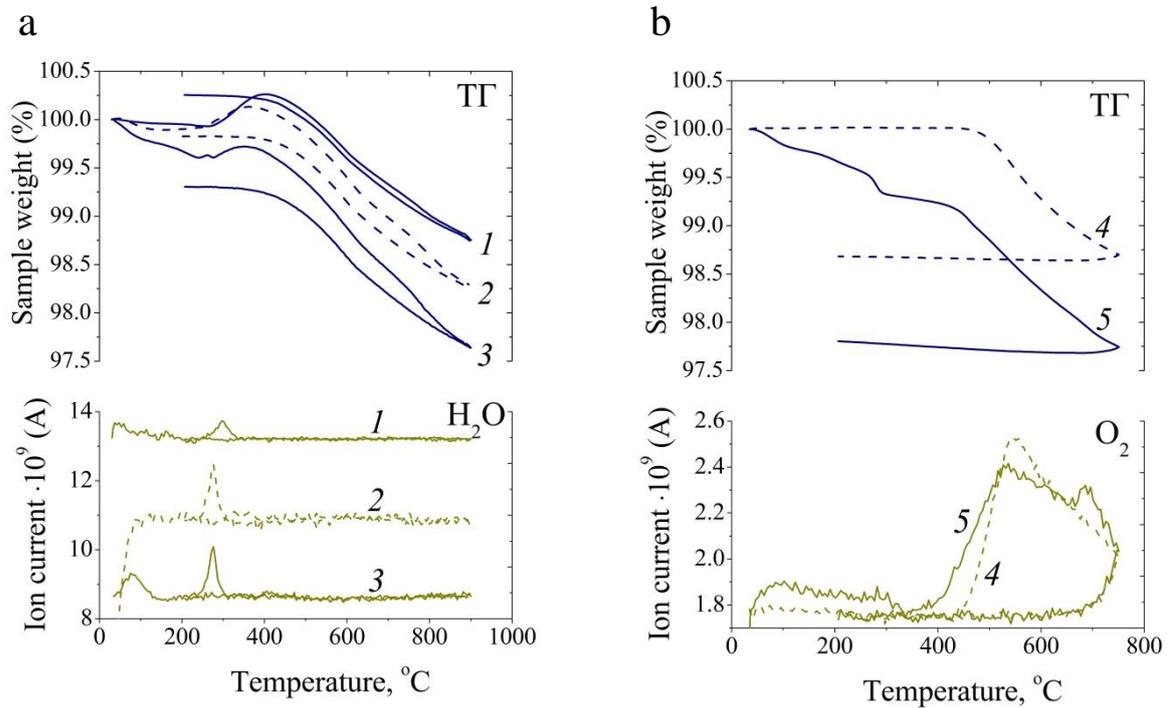

**Fig. 3.** (a) The TA results of the YBCO samples saturated with air components. Experiments were carried out under an atmosphere of 21% $O_2$ + 79% $N_2$ at a heating/cooling rate of 10 degree·min$^{-1}$. Curve numbers in the figure correspond to numbers of samples in Table 1. To eliminate crowding, curves *1* and *3* in the bottom of the figure are shifted along the vertical axis. (b) The TA results of the YBCO samples saturated with air components. Measurements were carried out under an atmosphere of argon at a heating/cooling rate of 10 degree·min$^{-1}$. Curve numbers in the figure correspond to numbers of samples in Table 1.

**Table 1**

The increase in weight of YBCO samples during exposure to air ($\Delta m_\Sigma$) compared with the amount of water ($\Delta m_{H_2O}$) and oxygen ($\Delta m_{O_2}$) degassed from the samples in the course of subsequent TA.

| Sample No | $\Delta m_\Sigma$, wt.% | $\Delta m_{H_2O}$, wt.% | $\Delta m_{O_2}$, wt.% |
|---|---|---|---|
| 1 | 0.20 | 0.06 | – |
| 2 | 0.55 | 0.12 | – |
| 3 | 0.90 | 0.26 | – |
| 4 | 0 | 0 | 1.37 |
| 5 | 1.29 | 0.47 | 1.74 |

As for the residual substance involved in the RT-gas-exchange (with the proportion of 76±5%), additional thermal studies were carried out under argon atmosphere for it to be detected, Fig. 3b. Appropriated atmosphere was used in order to obtain acceptable signal intensities for oxygen and nitrogen. As shown in the experiment the intensity of the oxygen signal from the sample previously exposed to air is slightly higher than one from the initial sample (data are listed in Table 1). Signal from nitrogen is absent for both samples.

Fig. 4 shows a full set of TA data obtained in the present study. One can see that the total amount of gas released from YBCO is still significantly less than that expected according to $\Delta m_\Sigma$-values, even taking $O_2$ into account. Since nothing except water and oxygen was found with the aid of mass spectrometry, a sector at the top of the graph was named "Nothing".

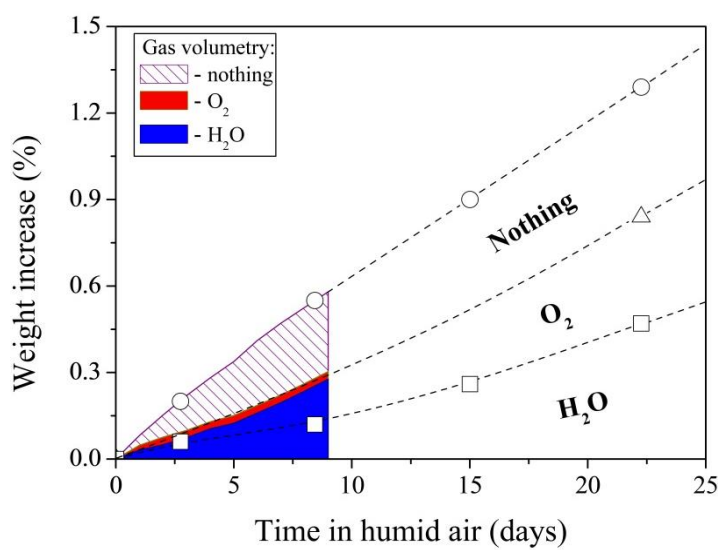

**Fig. 4.** The graphical representation of data concerning TA (signs) and gas volumetry (shaded areas). Circles, squares, and triangles are data of $\Delta m_\Sigma$, $\Delta m_{H_2O}$, and $\Delta m_{H_2O}+\Delta m_{O_2}$, respectively.

Hoping to get a concordance between $\Delta m_\Sigma$ and any other parameter characterizing mass transfer, we conducted a volumetric study of the type II samples when saturated with air components. For this purpose, we used the system of communicating vessels: a thermostatic mini chamber with a sample[1] and a horizontal transparent tube. The tube contained a small cylindrical glycerol mark that moved under the influence of pressure difference along the tube and fixed the amount of gas absorbed by a sample. Changing the atmospheric pressure was accounted by means of the appropriate correction term. Every 24 h a sample was short-term removed from the system to be weighted. In this experimental version, the total volume of gas absorbed by the sample ($V_\Sigma$) was found. In order to extract from $V_\Sigma$ the amount of water (i.e. to get the $V_{\Sigma-H_2O}$ quantity), in the other similar experiment, moisture holding material was placed in the sample vessel. Taking into account the TA results obtained previously, we accepted that $V_{\Sigma-H_2O} \equiv V_{O_2}$. Completed results of volumetric analysis are shown in Fig. 4 together with the data of TA.

First of all, the conspicuous incoordination between two different methods concerning $H_2O$ and $O_2$ data sets should be discussed – as we can see, TA shows much less water but more oxygen compared with volumetry. We consider the missing part of $H_2O$ in TA to be beyond the "visible" region of the $H_2O$ spectrum. This is consistent with the fact that YBCO was partly hydrolyzed during exposure to air (see Fig. 2). In this case, barium hydroxide with decomposition temperature of about 1000 °C must be formed; it is often not visible on X-ray spectra of YBCO. Indeed, taking into account the hydrolysis reaction [8], the amount of "missing" water in TA (about 0.35 wt.%) approximately corresponds to the amount of decomposition products observed in Fig. 2: $BaCuO_{2+x}$ and $Ba_2Cu_3O_{5+x}$ (about 2 vol.% each).

As for oxygen, its large quantity outflowing from the hydrated sample in the course of heating to 750 °C under an argon atmosphere (see Fig. 3b) is likely to result from an evident shift of this oxygen spectrum towards lower temperatures. In this case, the excess spectral weight of $O_2$ might come from the region $T > 750$ °C. As a proof of that, iodometric titration carried out on the initial and the exposed to air samples (Nos *4* and *5*, respectively) did not show any significant difference in the degree of oxidation of copper ions (δ values based this analysis are 0.75±0.02 and 0.77±0.02, respectively).

Meanwhile, the results of volumetry as well as the TA data indicate that a certain share of $\Delta m_\Sigma$ is not the result of mass transfer between the gas phase and YBCO, and there are no apparent reasons for its emergence. This situation is very similar to that described in the renowned paper [9], where mysterious weight changes in a $SiO_2$ sample suspended above a bulk YBCO superconductor were established.

---

[1] The chamber (volume of about 70 ml) was made from the standard stainless steel fittings and flanges of the ConFlat type designed for ultrahigh vacuum applications. Copper gaskets were used for its sealing. In order to avoid outgassing of the galvanic Fe-Cu couple, the chamber was separated from water of the thermostat by a plastic sheath.

Further, we describe the results of additional experiments that provide new information on the effect discovered here. In one case, we investigated the ability of two samples of type I to be saturated with gas together. Herewith, the overlapped area of the vertical projections of the samples was varied – by changing their inclination relative to the vertical, Fig. 5a. It was found that the overlapping of samples (a case of pairs *3*, *3'*; *4*, *4'* and *5*, *5'*), contrary to its apparent meaninglessness, would lead to changes in sample ability to be saturated with gas. Herewith a saturation rate of the upper samples turns out to be greater than that of the bottom samples, with the rate difference being dependent on difference in weight of the samples (see the caption of Fig. 5a). In the other case, it was observed that with weight of a type II sample decreasing from 800 to 250 mg, at a certain threshold of 500 mg ($W_{thr}$) YBCO suddenly ceased to absorb the gas, Fig. 5b. In this figure, instead of the weight, the height of the bulk layer of samples (h) is represented, which is related to the weight by an equation: $h[mm] = 7.3 \cdot 10^{-3} (W[mg] + 69)$. For the type I samples the study like this was not purposely conducted. However, for two samples with the smallest weight (W = 77 and 36 mg; h = 3 and 1 mm, respectively), which participated in our experiments, it was noted that the first one was saturated with gas with normal intensity yet, but the weight of the second one remained almost constant with time.

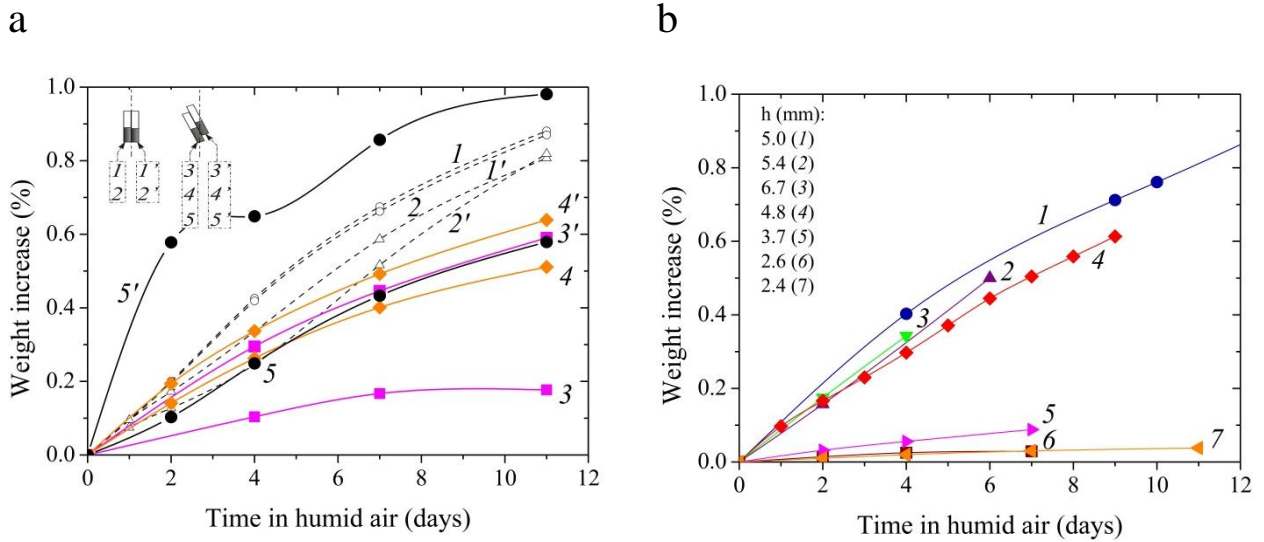

**Fig. 5.** (a) The influence of the sample position in space on the saturation of YBCO with the air components. The weight of the samples: *1* – 114 mg, *1'* – 149 mg; *2* – 146 mg, *2'* – 98 mg; *3* – 96 mg, *3'* – 139 mg; *4* – 131 mg, *4'* – 139 mg; *5* – 174 mg, *5'* – 77 mg; (b) The influence of the size h on the saturation of YBCO with the air components.

Additionally it has been established that the presence of electrostatic and magnetic fields does not affect the samples saturation up to $H = 1 \cdot 10^3$ Oe and $E = 50$ kV·m$^{-1}$, Fig. 6. However, when exceeding the electric field strength of 50 kV·m$^{-1}$ a sharp drop in the intensity of saturation occurs at once.

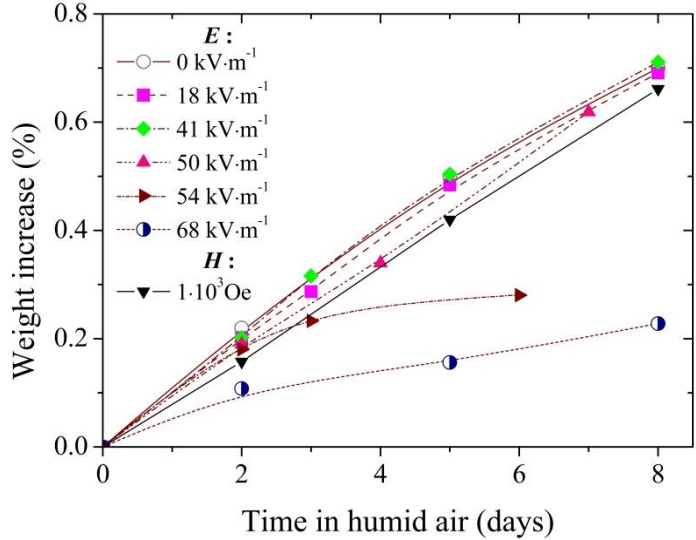

**Fig. 6.** The influence of the horizontal electric and magnetic fields on the saturation of YBCO with the air components.

At the moment, there is no basis for serious discussion of the nature of the observed new phenomenon. But with some confidence we can already say that the vertical impact on our samples described above is in fact only a particular manifestation of an attractive force (AF) of more general character originating in YBCO. With this one can explain the intense saturation of YBCO with water (and oxygen) at low $p_{H_2O}$ as well as the interaction of particles of the YBCO powder with each other, and the vertical force, which is evidently exerted by Earth.

We note that the results described in [9] can also be explained from this perspective.

Furthermore, the experimental results reflected in Figs 5 and 6 make it possible to form an approximate idea of the behavior and properties of AF. First of all, based on the available data, we can assume that a source of AF (AF center) has somewhat of force line passing through it. Besides that, this center can change its spatial orientation. Here, in our opinion, a constrained magnetic centers analogy of the AF centers would be appropriate. Particles of YBCO interacting with each other could be represented as interacting "domains", with a hard-ordered structure of AF lines crossing domain bulk. But domains themselves are ordered only under some external influence. The influence of the Earth on YBCO make AF lines be ordered vertically. Herewith, force lines come out on the open surface of a sample, acting on the gas phase components. Meanwhile, as shown in Fig. 6, the electric field can also interact with the AF lines. At a certain value of $E$ this interaction overpowers the influence of the Earth, reorienting AF lines horizontally. In this case, the saturation of YBCO with gas is terminated simply because of the closedness of the "active" surface of a sample by the walls of the tube. As for the threshold height of the bulk layer (see Fig. 5b), its existence may be explained by the aspiration of the system to a minimum quantity of AF lines in order to reduce the internal energy. That can be achieved by reducing the area of the outer sample surface that is crossed by AF lines by means of the latter being turned to horizontal plane, Fig. 7 (note that the threshold h ≈ 4 mm is the condition of the area equality of the base and the lateral surface of the cylindrical sample of type II).

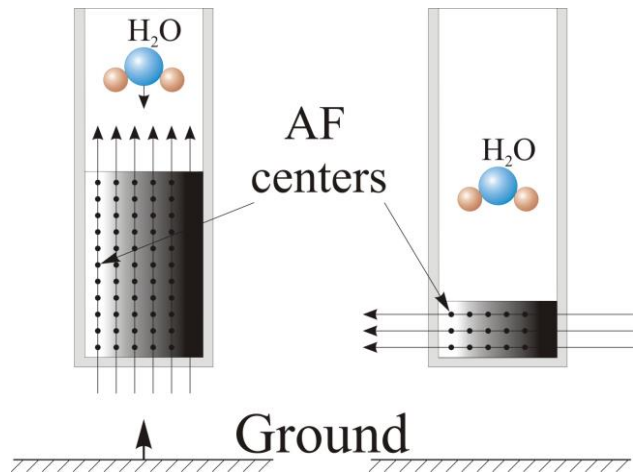

**Fig. 7.** Energetically favorable orientation of the AF lines in the cases of large and small samples of YBCO. When the AF lines are horizontal (see the right figure), the interaction of YBCO with the air components as well as with the Earth does not occur.

In accordance with the above position, the location of two YBCO samples one above the other when exposed to air should lead to the bilateral (multi-directional) impact of AF on the $H_2O$ ($O_2$) molecules in reaction space of the lower sample and to the unilateral AF influence in reaction space of the upper one. In the first case, the interaction between YBCO and gas would be weakened, in the second – enhanced. And then, the ratio of the weights of the two samples has to play a key role in the kinetics of their saturation with gas. This is in good agreement with the results of our experiments represented in Fig. 5a.

Fig. 8 shows results of the studies of magnetic properties of YBCO (samples *3* and *4* according to Table 1). It has been found that the magnetization of sample *4*, which not preliminarily exposed to air, begins sharply to fall when cooled below a characteristic critical temperature ($T_c$) of about 70 K, i.e., the sample behaves like a "regular" YBCO superconductor with $\delta = 0.75$ [10, 11]. Magnetic hysteresis loop shown in the inset of Fig. 8 reflects the typical processes going on in type-II superconductors: eddy current screening of the growing external magnetic field (there is a sharp inflection of the dependence in the area of the first critical field, $H_{c1} = 1866$ Oe); trapping and pinning of the magnetic flux with decreasing field strength. In turn, exposed to air YBCO represents a completely different behavior. The magnetization of sample *3* surprisingly increases when one is cooled below $T_c$ that is observed in fields of $0.05·10^4$ as well as of $1·10^4$ Oe. This increase occurs synchronized with the fall of the magnetization of sample *4*.

It should be noted that similar magnetic behavior was described in a number of works (for example, [12, 13]). That was observed for various types of superconductors and called the paramagnetic Meissner effect (PME). It is in general believed that PME arises under a very small external fields (about 1–10 Oe). Some exceptions, however, are known. For instance, in [13], the melt-textured YBCO sample with the addition of $Y_2BaCuO_5$ particles exhibits PME in fields up to $1.4·10^5$ Oe (this and similar effects are named High-Fields PME or HFPME). The authors of the latter work relate the appearance of HFPME to strong flux compression within weak or non-superconducting regions of the samples (in this case, the condition of the YBCO superconductor corresponds to the right-upper quadrant of the hysteresis loop, see the inset in Fig. 8). The phenomenon is considered to arise as a consequence of Meissner effect and assisted by strong flux pinning by a complex and correlated microstructure of the impurity particles.

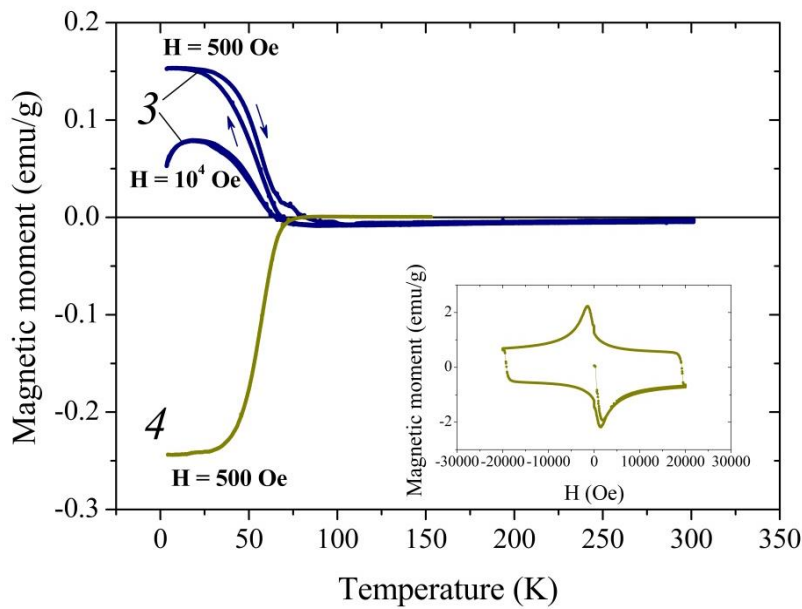

**Fig. 8.** The temperature dependence of the magnetization of the YBCO samples measured in the in-field cooling (FC) and in-field cooled warming (FCW) modes before and after exposure in air (curve numbers in the figure correspond to numbers of samples in Table 1). The inset shows the dependence of the magnetization of sample *4* on external magnetic field measured at 4 K.

As regards our case, there are reasons to link HFPME with the ability of the YBCO-matrix to affect various elements by AF. As a result, both the orientation of these elements in space and some kind of their ordering could take place.

On the other hand, if HFPME arises at a certain stage of development of the Meissner effect when YBCO is cooled, then it turns out that the long "tail" of negative magnetization observed for sample *3* up to 300 K (at 300 K the magnetic moment is $-0.73·10^{-3}$ emu under an external field of 500 Oe and $-1.34·10^{-3}$ emu under a field of $1·10^4$ Oe) is a manifestation of this effect. We explain such a high value of $T_c$ by AF centers possessing a specific property. According to Fig. 6, strong electric field (in particular, it can be from a charged particle in the YBCO structure) can reorient a AF center resulting in disturbance in the array of AF centers that in certain circumstances could cause the correlated interactions of pairs of electrons (holes) with it.

To confirm that the negative magnetization of the sample *3* was not an artifact and a kind of diamagnetism that would be expected in a perfect electrical conductor, we measured the weight of a large number of hydrated YBCO samples suspended above the source of controlled magnetic field (DC mode), as which we used an electromagnet (the magnetic field strength in the sample region was about $2 \cdot 10^3$ Oe; field gradient was not measured). Every time when the electromagnet was switched on, the sample weight was decreased by 0.5–2 wt.%.

## 4. Conclusion

Thus, in the paper it has been shown that exposure of YBCO to an air atmosphere at first at $p_{H_2O}$ = 110 Pa and then at $p_{H_2O}$ = 1 kPa leads to developing uncharacteristic and rather unusual properties in it. In particular, it has been observed the extremely excessive weight gain of YBCO samples that was not corresponding to the quantity of the absorbed gas as well as the Meissner effect at 300 K. All the results have been explained on the basis of an approximate idea that inside YBCO there is a certain source of attractive force (AF centers) acting on surrounding particles and bodies. Besides, AF centers can be influenced by the electric field of charged particles that result in disturbance in the array of AF centers and in certain circumstances could cause the correlated interactions of pairs of electrons (holes) with it.

**Acknowledgements**

We are grateful to collaborators of our laboratory for the help, particularly to S.Kh. Estemirova for X-ray analysis, S.A. Uporov and G.A. Dorogina for providing magnetic dependencies, and L.A. Cherepanova for the testing our samples by the iodometric titration method. The study was performed with the use of equipment of the Ural-M Collective Use Center at the Institute of Metallurgy of the Russian Academy of Sciences.